\documentclass[aps,pra,twocolumn]{revtex4-1}

\usepackage{amsfonts}
\usepackage{graphicx}
\usepackage{amsmath}
\usepackage{color}
\usepackage{braket}

\begin{document}
	
\title{Nonlinear Graphene Quantum Capacitors \textcolor{black}{for Electro-optics}}
	
\author{Sina Khorasani}
\affiliation{School of Electrical Engineering, Sharif University of Technology, Tehran, Iran}	
\affiliation{\'{E}cole Polytechnique F\'{e}d\'{e}rale de Lausanne, CH-1015, Lausanne, Switzerland}
\email{khorasani@sharif.edu}
\email{sina.khorasani@epfl.ch}
\author{Akshay Koottandavida}
\affiliation{UM-DAE Centre for Excellence in Basic Sciences, Vidyanagari, Mumbai, 400098, India}	
\affiliation{\'{E}cole Polytechnique F\'{e}d\'{e}rale de Lausanne, CH-1015, Lausanne, Switzerland}

\begin{abstract}
Owing to its peculiar energy dispersion, the quantum capacitance property of graphene can be exploited in a two-dimensional layered capacitor configuration. \textcolor{black}{Using graphene and boron nitride respectively as the electrodes and the insulating dielectric,} a strongly nonlinear behavior at zero bias and small voltages \textcolor{black}{is obtained}. When the temperature is sufficiently low, the strong nonlinear interaction emerging from the quantum capacitance \textcolor{black}{exhibits} a diverse range of phenomena. \textcolor{black}{The} proposed structure could take over the functionalities of nonlinear elements in many cryogenic quantum systems, and in particular, quantum electro-optics. It is shown that ultrastrong coupling \textcolor{black}{is easily reached} with small number of pump photons at temperatures around 1K and capacitor areas of the order of $1\mu{\textrm{m}}^2$. \textcolor{black}{A measure of anharmonicity is defined and as} potential applications, a qubit design as well as schemes for non-reciprocal devices such as an electromagnetic frequency circulator \textcolor{black}{are discussed}.
\end{abstract}


\maketitle

\section*{Introduction}

In the \textcolor{black}{fields} of quantum optomechanics \cite{1,2,2a,3,4} and superconducting circuits, it is desirable to have a nonlinear interaction between modes \textcolor{black}{with} different frequencies. For this purpose, optomechanical crystals \cite{1,2,4a}, nanomechanical capacitors \cite{2,2a,5}, kinetic inductors \cite{6,7}, and Josephson junctions \cite{8,9,10,11,12} are being \textcolor{black}{explored}. \textcolor{black}{By contrast,} in superconductive quantum computing \cite{12a,12b,12c}, circuit analogues of quadratic optomechanics \cite{12d}, and more recently realization of meminductors and memcapacitors \cite{12e} have been proposed. 

While nanomechanical capacitors effectively appear as third-order nonlinear Kerr elements in the circuits, Josephson junctions, depending on the structure fabrication type and magnetic flux bias, may exhibit various degrees of \textcolor{black}{second or third order} nonlinear flux/current behavior \cite{12f}. Josephson junctions, despite their strong nonlinearity, normally operate at very small currents, which limits their maximum total photon occupation number. \textcolor{black}{Moreover, they occasionally need to be biased by either a DC magnetic field or electric current.} As for the nonlinear capacitors, Graphene has been used in optomechanics as the mechanical element in the form of a drum capacitor \cite{15,16,17}. 

Here, we suggest an \textcolor{black}{alternative} solution by proposing the design of a nonlinear quantum capacitance. The capacitor is formed by a graphene-boron nitride-graphene layered two-dimensional (2D) structure\textcolor{black}{. This stacked 2D structure} has been extensively studied for its various electronic and optoelectronic properties \cite{13,14}. \textcolor{black}{This} particular design avoids physical mechanical motion as the structure is not suspended. In fact, this provides an alternative design for \textcolor{black}{non-dissipative} microwave quantum electro-optics \cite{18,19,20}. 

\textcolor{black}{It is required that the} Boron nitride (BN), or any other \textcolor{black}{substitute} wide-band-gap 2D material, be at least a few monolayers thick to prevent tunnel currents. \textcolor{black}{This is to avoid degrading} the performance of the quantum capacitor. The total electrical property of this layered structure is then determined by a geometric parallel-plate capacitor, placed in series with the graphene quantum capacitor. The quantum capacitance at zero bias and low temperatures takes on very small values, which overrides the geometric capacitance. Hence, the nonlinear quantum capacitance would dominate the overall behavior at sufficiently low temperatures and zero or small bias voltages. \textcolor{black}{Electronic properties of graphene \cite{20a,20b}, stemming from its Dirac-cone shaped band structure are behind this quantum capacitance effect \cite{20b1,20b2}. In fact, the nonlinear property of this quantum effect has not been unnoticed and used by other authors for different applications such as sensors \cite{20b3,20b4}. Quantum capacitance has been observed in Cooper-pair transistors \cite{20c} and boxes \cite{20d}, as well as qubit resonantor \cite{20e} and Nano-Electro-Mechanical Systems (NEMS) \cite{20f}.}

It is shown that \textcolor{black}{the} strongly nonlinear feature of quantum capacitance may be well exploited to obtain an interaction Hamiltonian between multiple electromagnetic modes, with an adjustable interaction rate through electrostatic gating, all the way down to the temperatures around zero. The existence of such a nonlinear capacitance element compatible with low temperatures and quantum regimes, may significantly contribute to the design flexibility and the application range of superconducting circuit \cite{12a,12b,12c}, circuit quantum cavity eletrodynamics, as well as quantum circuits and networks \cite{20g}. 

\section{Theory}

\subsection*{Quantum Capacitance}
\begin{figure}
	\centering
	\includegraphics[width=\linewidth]{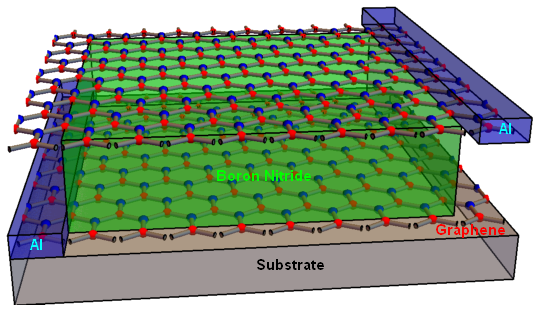}
	\caption{Schematics of the proposed layered structure.}
	\label{fig:2DCap}
\end{figure}

The per-unit capacitance of the 2D layered structure shown in Fig. \ref{fig:2DCap} is composed of two components, which arise from geometric and quantum properties. These two appear in series as $C=C_G||C_Q$, where $C_G=\epsilon/t$ is the geometric capacitance per unit area with $\epsilon$ and $t$ being the permittivity and thickness of the BN layer. Also, \textcolor{black}{$C_Q=\frac{d}{dV}Q$}, is the quantum capacitance per unit area of the graphene electrode pair given by \cite{21,22,23}
\begin{equation}\label{eq1}
  C_Q=\frac{2e^2k_B T}{\pi (\hbar v_F)^2}\textrm{ln}\left[2\left(1+\cosh\frac{E_F}{k_B T}\right)\right],
\end{equation}
\noindent
where $e$ is the electronic charge, $k_B T$ is the thermal energy, $v_F\approx c/300$ is the Fermi velocity of graphene, and $E_F=eV/2$ is the Fermi energy at either of the graphene electrodes, with $V$ being the voltage drop across $C_Q$. In general, $V$ is not the same as the external applied voltage $V_{\rm ext}$, but under condition $C_G>>C_Q$ discussed below, it is reasonable to ignore this difference and thus let $V\approx V_{\rm ext}$. \textcolor{black}{It should be however mentioned that achieving this regime might be not quite trivial due to disorder and charge impurities which induce electron-hole puddles \cite{23x}. Furthermore, at very small $E_F$, the Fermi velocity $v_F$ in graphene should be renormalized due to electron-electron interactions, leading to further extra logarithmic corrections \cite{23y}. }

The typical values of $C_Q$ is in the range of $1-10{\rm fF\mu m^{-2}}$ \cite{23}. In the ultralow temperature limit, the above expression approaches the linear form
\begin{equation}\label{eq2}
  \textrm{lim}_{T\rightarrow0} C_Q=\frac{e^3}{\pi (\hbar v_F)^2}|V|.
\end{equation}
\noindent
Hence, as the extreme limit for the symmetric structure under consideration with $E_F=0$ at zero bias and zero temperature, the differential capacitance vanishes $C_Q=0$. Though not being practical, this limit underlines the fact that the overall capacitance of the structure at zero bias, sufficiently low temperatures, and low drive-voltage is solely determined by its quantum properties. Although being a bit counter-intuitive, this is exactly what one would need to exploit the resulting nonlinear property of the quantum capacitor. In what follows, we explore the limits of this dominance, and calculate the expected nonlinear interactions in details.

\subsection*{Series Expansion}

At zero temperature, following (\ref{eq2}), we obtain the nonlinear charge-voltage relationships
\begin{eqnarray}\label{eq3}
  Q&=&\int_{0}^{V}CdV=\frac{e^3}{2\pi (\hbar v_F)^2}|V|V,\\ \nonumber
  U&=&\int_{0}^{V}QdV=\frac{e^3}{6\pi (\hbar v_F)^2}|V|^3.
\end{eqnarray}
\noindent
where $U=\frac{1}{3}\sqrt{2\pi}\hbar v_F \textrm{sgn}(Q)N^\frac{3}{2}$ with $|Q|=eN$ is the total stored energy per unit area. These relations, despite their simplicity, are unfortunately both singular at $V=0$ due to the appearance of absolute value $|\cdot|$, which prohibits further expansion. Hence, (\ref{eq3}) cannot be used as a basis of field quantization.

\begin{figure}
	\centering
	\includegraphics[width=\linewidth]{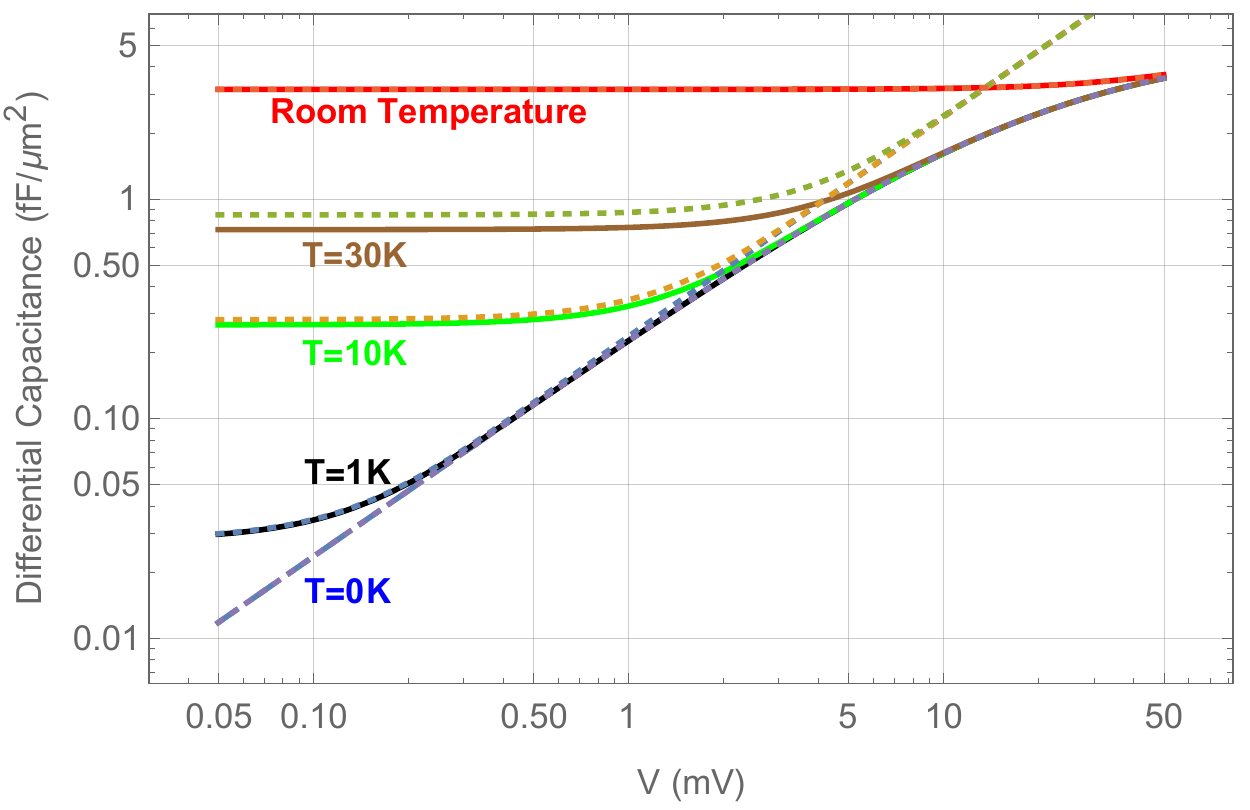}
	\caption{The differential capacitance of the proposed layered structure at various temperatures on the logarithmic scale. Dashed line is the zero temperature limit. Dotted lines correspond to the quantum capacitance $C_Q$ only, without taking the effect of geometric capacitance $C_G$ into account.}
	\label{fig:QuantumCap}
\end{figure}

However, it is possible to proceed directly from (\ref{eq1}) at finite temperature instead, which after considerable algebra, allows us to obtain the series expansions
\begin{eqnarray}\label{eq4}
N&\approx&\frac{2e k_B T}{\pi(\hbar v_F)^2}\left[\ln (2) V+\frac{e^2}{96 (k_B T)^2}V^3+\dots\right],\\ \nonumber
U&\approx&\frac{\pi (\hbar v_F)^2}{2k_B T}\left\{\frac{N^2}{\ln(16)}-\frac{\pi^2}{4}\left[\frac{\hbar v_F}{\ln(16)k_B T}\right]^4N^4+\dots\right\}.
\end{eqnarray}
\noindent
Firstly, it is seen from the first expansion term of the second equation, that the linear capacitance per unit area is given as
\begin{equation}\label{eq4a}
C_0=\frac{2e^2k_B T \ln(16)}{\pi (\hbar v_F)^2}.
\end{equation}
\noindent
This is to be compared with the geometrical capacitance $C_G$. The permittivity of BN is $\epsilon=4.0\epsilon_0$ \cite{13}. Hence taking the dielectric thickness to be $t=7\textrm{nm}$, being sufficient to stop tunneling current, gives a geometric capacitance value of $C_G=5.06 {\rm fF\mu m^{-2}}$. At the typical temperature of $1\textrm{K}$, we have $C_0=0.0563{\rm fF\mu m^{-2}}$, which is much smaller than the geometric capacitance by two orders of magnitude. Hence, our initial assertion to ignore the effect of the geometric capacitance at low temperatures and zero bias is satisfied to a high accuracy. This situation is illustrated in Fig. \ref{fig:QuantumCap}. 
In general, a dielectric thickness of no less than $2{\rm nm}$ should be sufficient to completely disallow tunneling. It could be made thick as long as $C_Q$ dominates by an order of magnitude, yielding the available range of dielectric thickness as $3{\rm nm}<t<70{\rm nm}$. Since, the monolayer BN is around only $2.489{\rm \AA}$ in thickness \cite{13}, $t$ is sufficiently high to warrant an expitaxial growth. Structures of comparable type are being fabricated in integrated optics \cite{23a,23b}. In general, there is no reason why other crystalline dielectrics such as ${\rm SiO_2}$ or ${\rm Al_2O_3}$ could not be used.

\textcolor{black}{
It is remarkable that the quantum capacitance (\ref{eq4a}) is inherently independent of the dielectric thickness. On one hand, a dielectric too thin would allow tunneling currents which appear as a parallel nonlinear resistance across the capacitor. This might be useful in some applications such as light emission \cite{23a,23b}, however it is highly non-desirable for this type of circuit element and must be avoided. Hence, a minimum dielectric thickness of roughly $3\textrm{nm}$ is needed to ensure that tunneling currents are fully blocked. On the other hand, a dielectric too thick would cause quantum capacitance to fade away while appearing in series with the geometric capacitance. The reason is that it is the geometric capacitance which decreases significantly at small dielectric thickness. Quantum regime is accessible only if the quantum capacitance is much smaller than the geometric capacitance. These criteria set the above mentioned boundaries on the minimum and maximum thickness of the insulating barrier.	
}

\subsection*{Quantized Hamiltonian}
Now, suppose that the capacitor is driven by a sinusoidal voltage with angular frequency $\omega$ and a fixed phase $\phi$. The corresponding charge density operator is given as
\begin{equation}\label{eq5}
  \hat{Q}=e\hat{N}=\frac{e\psi}{\sqrt{2}}(\hat{a}+\hat{a}^\dagger),
\end{equation}
\noindent
where $\hat{a}$ is the bosonic field operator, satisfying $[\hat{a},\hat{a}^\dagger]=1$, and $\psi$ is the fluctuation amplitude of corresponding single-photon number density. Hence, $\psi$ is obtained from (\ref{eq4}) as
\begin{equation}\label{eq6}
  \psi=\sqrt{\frac{k_B T\ln(16)}{2\pi S\hbar v_F^2}\omega}=\chi\sqrt{\omega}.
\end{equation}
\noindent
Here, $S$ is the total capacitor area. After adding a resonant inductive term to the Hamiltonian as $\frac{1}{2}L\dot{\hat{q}}^2$ with $\hat{q}=S\hat{Q}$ being the operator of total charge on the 2D quantum capacitor and $\omega=1/\sqrt{SLC_0}$, this allows us to indicate the Hamiltonian of such a sinusoidal field from (\ref{eq4}) as
\begin{equation}\label{eq7}
  \hat{H}=\hbar\omega\left(\hat{a}^\dagger\hat{a}+\frac{1}{2}\right)
  -\frac{\hbar\tau}{4}\omega^2(\hat{a}+\hat{a}^\dagger)^4,
\end{equation}
\noindent
where
\begin{equation}\label{eq8}
  \tau=\frac{\pi^3 S\hbar^5 v_F^6\chi^4}{2\ln^4(16) (k_B T)^5}=\frac{\pi \hbar^3 v_F^2}{8\ln^2(16)S(k_B T)^3},
\end{equation}
\noindent
is a characteristic time constant of nonlinear interaction. Clearly, (\ref{eq7}) represents an anharmonic oscillator \cite{12a,12b,12c} and admits a Kerr-like nonlinearity, and this compares well to that of the Josephson junctions \cite{8,9,10,11,12}. It is not difficult to verify that the range of validity of the anharmonic approximation (\ref{eq8}) is limited to $\hbar\omega\approx e|V|<2k_B T$, which by expressing $T$ in ${\rm K}$ and frequency in ${\rm GHz}$ gives the upper photon number limit of $n_{\rm max}=41.7 T/f$. Hence, at $T=1{\rm K}$ and frequencies in the ${\rm GHz}$ range, many photons may interact anharmonically. Beyond this value, the 6th order nonlinearity kicks in, which remains valid in the range $4k_B T<e|V|<6k_B T$, and so on for higher order interactions \cite{23b2}. Similarly, at sub-Kelvin temperatures, the usefulness of this nonlinear capacitor may become limited to sub-GHz photons. It should be stressed out that one could access a 3-rd order interaction according to (\ref{eq3}) via appropriate DC voltage bias.

It is clear from the above that the nonlinearity is pronounced at small capacitor areas and low temperature. We shall observe that at the typical GHz microwave frequencies and temperatures below $1\textrm{K}$ this expression easily gives rise to a quite significant Kerr interaction.

Next, we may define a measure of anharmonicity for applications in qubit design \cite{23c} as $A=|1-(\omega_{21}/\omega_{10})|\times 100\%$, where $\omega_{mn}$ is the transition frequency between levels $m$ and $n$. A straightforward calculation gives the estimate $A=12\tau\omega$. Putting $T$ in ${\rm K}$, $f$ in ${\rm GHz}$, and $S$ in $\mu\textrm{m}^2$ gives $A=42.85\% f/S T^3$. Using the typical values $T=1\textrm{K}$, $f=4\textrm{GHz}$, and $S=100\mu\textrm{m}^2$ gives an anharmonicity of $A=1.1714\%$, while at $T=0.5\textrm{K}$ we obtain $A=13.71\%$, being sufficiently large for qubit applications.

\textcolor{black}{As long as the first $0\rightarrow 1$ and second $1\rightarrow 2$  transitions are sufficiently separated in frequency, one could ensure that only exactly one photon with frequency $\omega_{01}$ could exist in the system. This is how an ideal qubit should operate \cite{12a,12b,12c}, since only the photon states $\ket{0}$ and $\ket{1}$ could be excited by interaction with a radiation reservoir at frequency $\omega_{01}$. To achieve this, the anharmonicity $A$ should be only much larger than the normalized linewidth of the radiation $\Delta\omega/\omega_{01}$. Practically, the latter is less than a hundredth of a percent, so an anharmonicity of $A=13.7\%$ is quite large. As a result, creation of superposition states $\ket{\alpha}=c_0\ket{0}+c_1\ket{1}$ with $c_j,j=1,2$ being complex constants and $\braket{\alpha|\alpha}=1$, becomes possible.}

\textcolor{black}{	
By setting a parallel linear resonant inductor and an optional series linear capacitor, this lays down the foundation of a novel type of qubit based on nonlinear quantum capacitance without need to external magnetic bias. This is to be compared with other types of nonlinear inductances, such as Josephson junctions \cite{8,9,10,11,12} and kinetic inductance \cite{6,7}. Since, a GHz microwave qubit should typically operate at cryogenic temperatures, the area of the quantum capacitance should probably be increased to avoid overly strong interaction. It remains, however, a matter of speculation that whether this proposed type of qubits could sustain their coherence over longer time intervals comparing to other superconducting technologies.}

\subsection*{Multi-mode Fields}
In a similar manner, for a multi-mode field composed of a few sinusoidal terms and by adding resonant inductive terms as $\frac{1}{2}\Sigma L_n\dot{\hat{q}}_n^2$ with $\hat{q}_n=S\hat{Q}_n$ and $\omega_n=1/\sqrt{SL_n C_0}$, we may write down the total Hamiltonian
\begin{equation}\label{eq9}
  \hat{H}=\hbar\sum_{n}\omega_n\left(\hat{a}_n^\dagger\hat{a}_n+\frac{1}{2}\right)
  -\frac{\hbar\tau}{4}\left[\sum_{n}\sqrt{\omega_n}(\hat{a}_n+\hat{a}_n^\dagger)\right]^4.
\end{equation}
\noindent
Here, we have the usual commutation rules between individual field operators as $[\hat{a}_n,\hat{a}_m^\dagger]=\delta_{nm}$ and $[\hat{a}_n,\hat{a}_m]=0$. Now, we limit the discussion to only three modes, where the interacting $\mathbb{H}$ and non-interacting $\hat{H}_0$ terms may be separated as $\hat{H}=\hat{H}_0-\mathbb{H}$ to yield the three basic types of interactions in the Hamiltonian
\begin{equation}\label{eq10}
\mathbb{H}=\mathbb{H}^{(1)}+\mathbb{H}^{(2)}+\mathbb{H}^{(3)},
\end{equation}
\noindent
with
\begin{eqnarray}\label{eq11}
\mathbb{H}^{(1)} &=& \frac{1}{2}\sum_{n \neq m}\hbar \gamma_{nmm}(\hat{a}_n+\hat{a}_n^\dagger)^2(\hat{a}_m+\hat{a}_m^\dagger)^2,\\ \nonumber
\mathbb{H}^{(2)} &=& \frac{3}{2}\sum_{n \neq m}\hbar \gamma_{nmm}(\hat{a}_n+\hat{a}_n^\dagger)^3(\hat{a}_m+\hat{a}_m^\dagger),\\ \nonumber
\mathbb{H}^{(3)} &=& \frac{3}{2}\sum_{n\neq m \neq l}\hbar \gamma_{nml}(\hat{a}_n+\hat{a}_n^\dagger)^2(\hat{a}_m+\hat{a}_m^\dagger)(\hat{a}_l+\hat{a}_l^\dagger).
\end{eqnarray}
\noindent
Here, $\gamma_{nml}=\tau \omega_n \sqrt{\omega_m \omega_l}$.

We can suppose that the interaction is driven at mode $0$ with a strong coherent field and frequency $\Omega=\omega_0$, in such a way that $\hat{a}_0\rightarrow \bar{a}+\delta\hat{a}$. Neglecting beyond simple nonlinear interactions between modes $1$ and $2$, and with three modes interacting, only $\mathbb{H}^{(3)}$ contributes significantly. Then, we obtain
\begin{equation}\label{eq12}
\mathbb{H} = \hbar \bar{G} (\hat{a}_1+\hat{a}_1^\dagger)(\hat{a}_2+\hat{a}_2^\dagger),
\end{equation}
\noindent
where $\bar{G}=3 \gamma_{012}(\bar{a}+\bar{a}^*)^2$.

Now we use the Rotating Wave Approximation (RWA) and neglect the counter-rotating terms, to obtain the remaining significant interactions. The RWA approximation is not valid in general, and in some cases, the counter-rotating terms have been shown to have a considerable effect on the evolution of system \cite{24,25}. Such a Hamiltonian has been achieved in systems using Josephson junction \cite{35} and extended to realise non-reciprocal devices \cite{34}. Anyhow, by application of RWA one may distinguish two particular resonant cases with $2\Omega=\omega_1\pm\omega_2$, for which the interaction Hamiltonian (\ref{eq10}) greatly simplifies and takes either of the forms
\begin{equation}\label{eq13}
\mathbb{H}\approx\hbar G (e^{2j\theta}\hat{a}_1\hat{a}_2^\dagger +e^{-2j\theta}\hat{a}_1^{\dagger}\hat{a}_2),
\end{equation}
\noindent
for $2\Omega=\omega_1-\omega_2$ and
\begin{equation}\label{eq14}
\mathbb{H}\approx\hbar G (e^{2j\theta}\hat{a}_1^{\dagger}\hat{a}_2^\dagger +e^{-2j\theta}\hat{a}_1\hat{a}_2),
\end{equation}
for $2\Omega=\omega_1+\omega_2$. Also, $G=g_0|\bar{a}|^2=3 \gamma_{012}|\bar{a}|^2$ is the overall interaction rate and $\theta=\angle\bar{a}$ is the phase of the coherent drive.

On one hand, by choice of $2\Omega=\omega_1-\omega_2$ and the Hamiltonian (\ref{eq13}), one may achieve a beam-splitter or hopping interaction, where the mode at frequency $\Omega$ is pumped strongly by a coherent field. This type of interaction can facilitate, for example, a non-reciprocal response between the modes under the appropriate conditions \cite{32,36}.

On the other hand, by choice of $2\Omega=\omega_1+\omega_2$ and the Hamiltonian (\ref{eq14}), one may achieve the parametric amplifier interaction. We can set up a system, similar to that discussed in \cite{34}, in order to realise a directional amplifier. However, the coupling is mediated by the quantum capacitor, as opposed to a Josephson junction, which in principle requires a  magnetic bias to induce the third order interaction in the system.

\textcolor{black}{Remarkably, this strategy to select the desirable type of interaction by appropriate selection of pump frequency has been verified experimentally in a recent experiment \cite{36a}, where non-reciprocal transmission of microwave signals is achieved in a nonlinear superconducting circuit based on ${\rm Nb/Al_2O_3/Nb}$ Josephson junctions with a biased external DC magnetic field. Thereby, a Field Programmable Josephson Amplifier (FPJA) is obtained which is capable of performing different functionalities. The clear possible advantage of the proposed nonlinear capacitor is no need to external magnetic field.}

\subsection*{Single-Photon Interaction}

It is possible to estimate the interaction from the above relations for $G$. Expressing the temperature $T$ in $\textrm{K}$, capacitor area $S$ in $\mu{\textrm{m}}^2$, and drive frequency $f=\Omega/2\pi$, electromagnetic frequencies $f_n=\omega_n/2\pi$ and also the single-photon interaction rate $g_0$ in $\textrm{GHz}$, we obtain
\begin{equation} \label{rate}
  g_0 =2\pi\times 0.143 \frac{f\sqrt{f_1 f_2}}{S T^3}.
\end{equation}
\noindent
As an example, taking $S=100{\rm \mu m^2}$ at $T=1{\rm K}$, with $f=4{\rm GHz}$, $f_1=2{\rm GHz}$, and $f_2=10{\rm GHz}$ satisfies the resonance condition for the hopping interaction (\ref{eq13}). This gives a linear capacitance value of $C_0=5.63{\rm fF}$ and also makes it possible to calculate the values of corresponding tank inductors and external circuitry. For these values, the single-photon interaction rate is $g_0=2\pi\times 25.55{\rm MHz}$. This fairly well falls into the domain of strong coupling. The same value at the temperature of $T=4{\rm K}$ is severely degraded to the weakly coupled value $g_0=2\pi\times 399.2{\rm kHz}$, while at the lower temperature of $T=0.25{\rm K}$, it is significantly enhanced to $g_0=2\pi\times 1.635{\rm GHz}$, which is of course ultrastrongly coupled. Hence, it is \textcolor{black}{possible} to obtain extremely strong single-photon nonlinear interaction rates even with low pump levels, without need to dilution refrigerators. \textcolor{black}{At lower temperatures, system enters the ultrastrongly-coupled regime and $g_0$ can even exceed the system resonant frequencies \cite{36b}, while at a higher temperature above a few Kelvins, the nonlinearity becomes too weak. This is a result of either (\ref{eq8}) or (\ref{rate}), where there is a $\propto S^{-1}T^{-3}$ dependence of the anharmonicity on area and temperature. Furthermore, the area $S$ could be freely chosen according to the operation frequency $f_{j},j=1,2$ and temperature $T$ to yield the desirable interaction strength $g_0$. Typically for GHz microwave circuits at liquid Helium temperatures, it could be in the range of $1-100{\rm \mu m^2}$.}

Other issues to be considered here is the superconductor contact to the graphene electrodes, as well as the charging time of the capacitor. The first issue brings up the proximity effect, while the second is related to the intrinsic conductivity of the graphene. The proximity effect has been well studied \cite{26,27,28,29}. It has been experimentally and theoretically established that in graphene, the finite normal state conductance can maintain finite supercurrent, even at zero bias. Hence, the electrical current should have no problem flowing freely through graphene electrodes. Moreover, it is well known that carriers in graphene are essentially ballistic \cite{29a,29b} at low temperatures over distances of a few microns.

\subsection*{Charging Delay}
Next, we may turn to the issue of charging delay. This issue has been studied in mesoscopic capacitors \cite{29c} and a universal charge resistance has been found for normal conductors. Remarkably, at zero bias, the steady state charge density of graphene is also expected to be zero, which should admit no normal state conductance, but it is well known also that the conductance of graphene at zero bias is also nonzero and finite, being in fact equal to the quantum conductance $\sigma_Q=2e^2/\pi\hbar$ \cite{29,30,31}.

In that sense, the ratio of quantum capacitance to quantum conductance with the unit of time, is simply proportional to the Fermi energy
\textcolor{black}{\begin{equation}\label{eq15}
  \tau_Q=S\frac{C_Q}{\sigma_Q}=S\frac{|E_F|}{\hbar v_F^2}.
\end{equation}}
\noindent
The time-constant $\tau_Q$ determines the quantum characteristic time for charge/discharge of capacitor when the current is supplied in the Ohmic regime. Not only for the structure under consideration, it approaches zero at zero bias, but also the electric current is mainly non-Ohmic and dissipation-free over distances of the order of few microns, since charges move almost ballistically across graphene electrodes. So, it should be safe to ignore both of these aspects in a precisely fabricated 2D layered structure, as long as it is void of defects and interfaces are atomically flat. \textcolor{black}{It must be pointed out that these restrictions on fabrication can be experimentally challenging to achieve, and are studied thoroughly in a recent article to which the readers are being referred \cite{36c}. Also, the minimal quantum conductivity $\sigma_Q$ may ideally not be achievable due to the disorder effects and finite temperature \cite{23y}.}

\textcolor{black}{Meanwhile, we note that for the quantum capacitor under consideration, the electrical current spreads horizontally over the graphene layers. Now since it is the area $S$ which is dominant, one may proportionally shrink the graphene length and increase the electrode width until the series resistance is negligible. If the length is sufficiently small, being less than mean free path of carriers at the cryogenic temperatures, then the expectation that the ballistic propagation of carriers would prevent Ohmic losses to appear, fully makes sense. It should be added that with the recent discovery of induced superconductivity in single-layer graphene \cite{36d} with a transition temperature of $4.2\textrm{K}$, hopes are on rise for a fully dissipation-free operation of the proposed system.}

\section{Microwave Circulator}
Now, as potential applications, we discuss schemes for two non-reciprocal devices. Assume that initially we have three electromagnetic modes with frequencies $\omega_1$, $\omega_2$, and $\omega_3$, mutually coupled via three nonlinear 2D capacitors, interacting through the Hamiltonian described in (\ref{eq13}). We set up the hopping interaction between each of the 3 modes using the scheme mentioned previously. Hence the Hamiltonian of the system could be written as $\mathbb{H}=\mathbb{H}_{12}+\mathbb{H}_{23}+\mathbb{H}_{31}$, where
\begin{eqnarray}\label{eq16}
\mathbb{H}_{12}&=&\hbar g_{3} (e^{i\phi_{3}}\hat{a}_1\hat{a}_2^\dagger+e^{-i\phi_{3}}\hat{a}_1^\dagger\hat{a}_2),\\ \nonumber
\mathbb{H}_{23}&=&\hbar g_{1} (e^{i\phi_{1}}\hat{a}_2\hat{a}_3^\dagger+e^{-i\phi_{1}}\hat{a}_2^\dagger\hat{a}_3),\\ \nonumber
\mathbb{H}_{31}&=&\hbar g_{2} (e^{i\phi_{2}}\hat{a}_3\hat{a}_1^\dagger+e^{-i\phi_{2}}\hat{a}_3^\dagger\hat{a}_1).
\end{eqnarray}
\noindent
It is worth mentioning here that the resonance condition for hopping interaction Hamiltonian, requires the constraint $\Omega_1+\Omega_3=\Omega_2$ on pump drives.

\begin{figure}
	\centering
	\includegraphics[width=\linewidth]{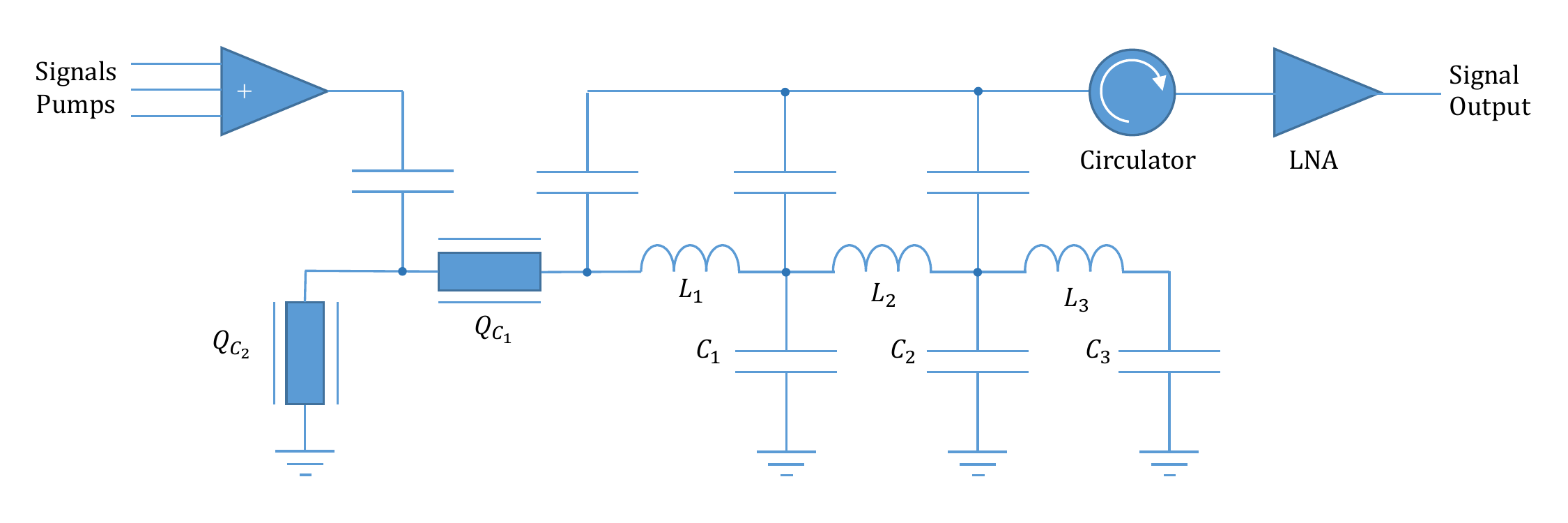}
	\caption{Proposed nonlinear circuit for the Field-Programmable Capacitor Array.}
	\label{fig:circuit}
\end{figure}

\textcolor{black}{
The corresponding circuit could be easily set up by making a topological dual equivalent circuit \cite{36e} of the Field-Programmable Josephson Array (FPJA) proposed recently elsewhere \cite{36a} for a superconducting non-dissipative reciprocal amplifier. This will lead to a circuit as shown in Fig. \ref{fig:circuit}, being here referred to as the Field-Programmable Capacitor Array (FPCA), held at cryogenic temperature. There are two nonlinear quantum capacitors $C_{Qj},j=1,2$, joined with six linear superconducting elements, composed of three linear tank inductors $L_j,j=1,2,3$ and three linear tank capacitors $C_j,j=1,2,3$. These determine the resonant frequencies of the system. Pumps feed through the quantum capacitors at desirable and fixed frequencies to achieve the needed interaction types and strengths. Collected signals are fed through a regular microwave circulator at a higher temperature to a Low Noise Amplifer (LNA), prior to detection and signal analysis. A similar strategy for a four-mode Diamond-shaped system could be implemented to obtain a non-dissipative microwave superconducting circuit, in order to achieve extremely high non-reciprocal amplification and isolation, while maintaining identical input/output frequencies \cite{36f}.} 

\begin{figure}
	\centering
	\includegraphics[width=\linewidth]{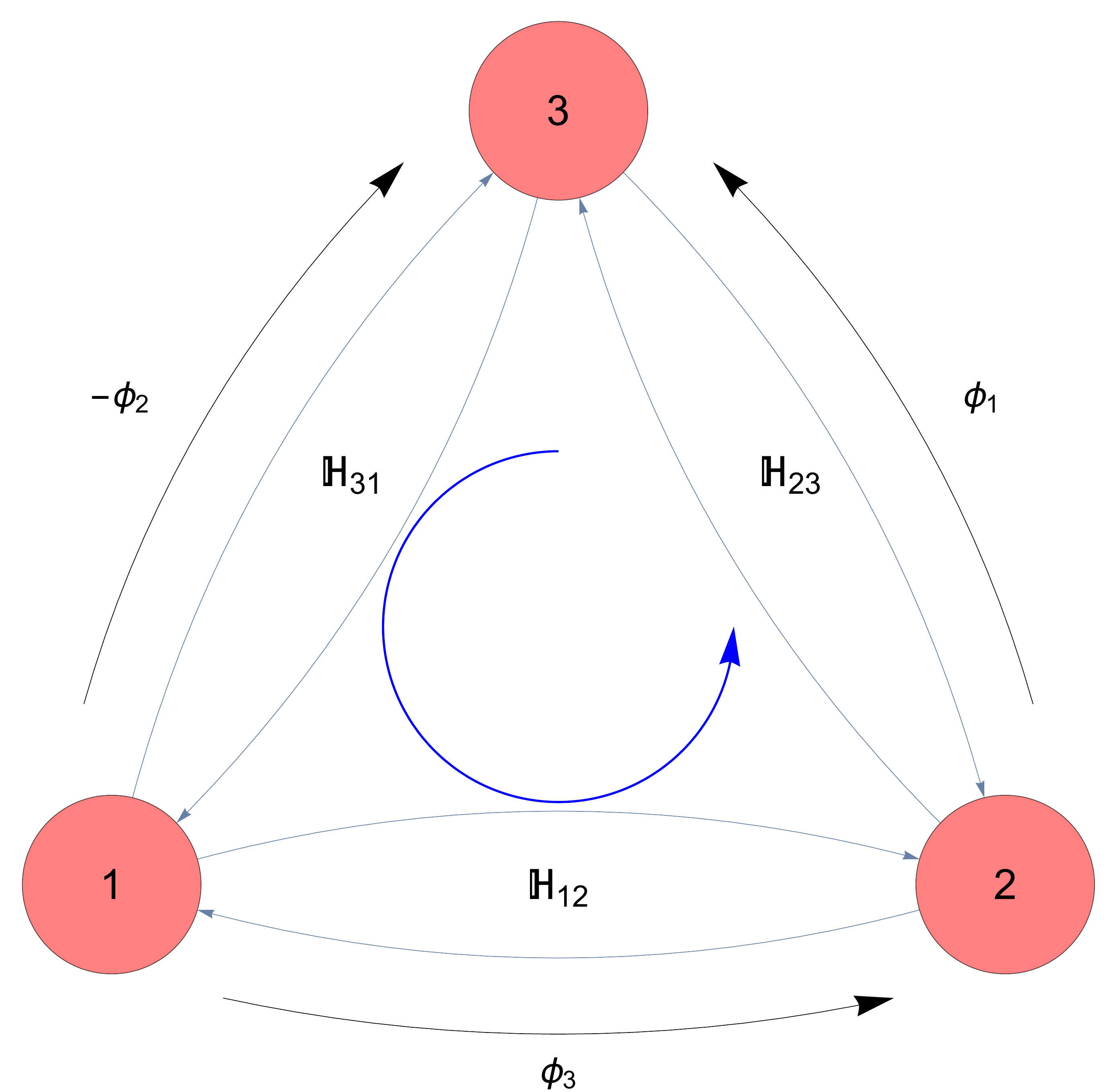}
	\caption{The mode coupling scheme in an ideal circulator. The signal can take two pathways for conversion from, say, mode 1 to mode 3. The phase difference between the two paths, $\Delta\phi$, can be controlled and by setting $\Delta\phi=\pm \frac{\pi}{2}$ provides the condition for interference between the paths. Hence only one pathway is preferred (say, anti-clockwise, blue curve) and not the opposite(clockwise) for mode conversion. In this scheme, any one of the 3 modes can be taken as the input while any of the other two may play the role of output.}
	\label{fig:circulator}
\end{figure}
\begin{figure}
	\centering
	\includegraphics[width=\linewidth]{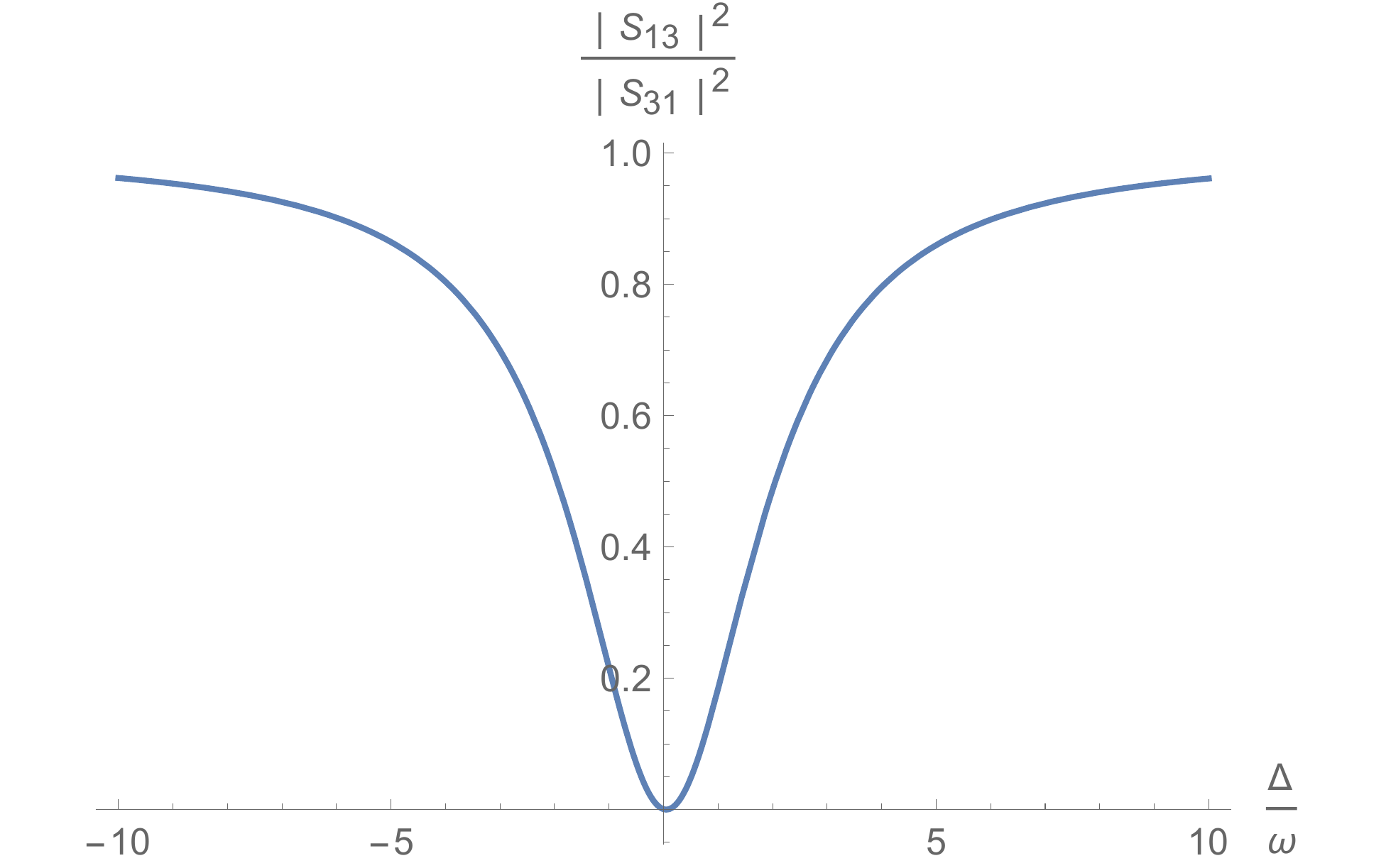}
	\caption{The ratio of between $1\rightarrow3$ and $3\rightarrow1$ mode conversion scattering amplitudes. Due to destructive interference, set up by the phase difference parameter, the symmetry of the system is broken.}
	\label{fig:sAmp}
\end{figure}
\begin{figure}
	\centering
	\includegraphics[width=\linewidth]{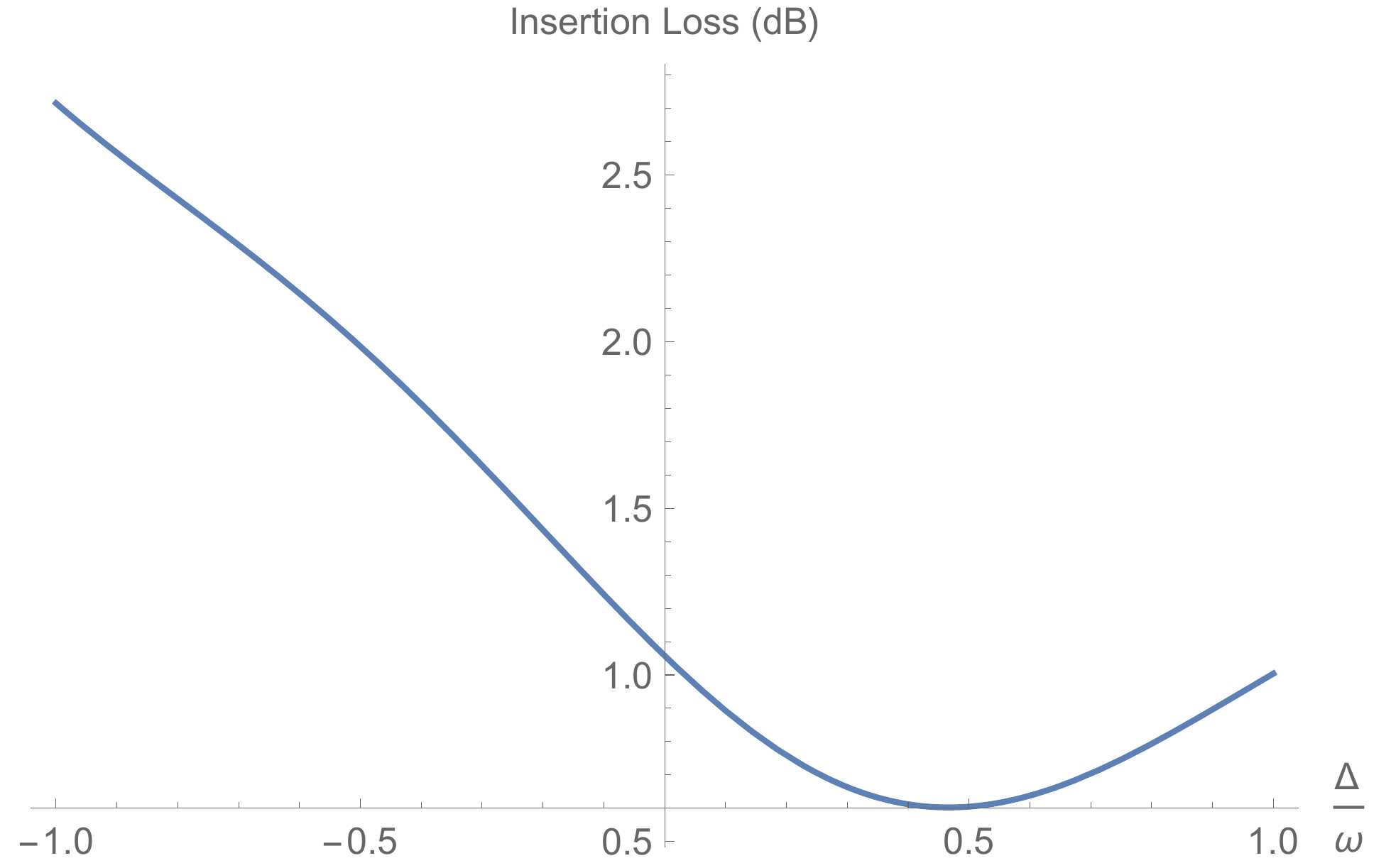}
	\caption{Illustration of insertion loss for the proposed circulator.}
	\label{fig:iLoss}
\end{figure}

A probe signal (input) at, say mode 1, has two pathways for conversion to mode 3 ($1\rightarrow2:2\rightarrow3$ or $1\rightarrow3$). The signal picks up different phases while traversing through these two paths \cite{33}. The phase difference at mode 3 (output) is $\Delta\phi =\phi_{1} +\phi_{3}-\phi_{2}$ (Fig. \ref{fig:circulator}). \textcolor{black}{Here, the horizontal axis is the normalized detuning $\Delta$ with respect to the resonant frequency $\omega$.} By controlling the phases of the pumps, one can set $\Delta\phi=\pm \frac{\pi}{2}$, thereby allowing interference between the two paths. This results in the signal propagating only in one of the two paths, hence breaking the symmetry of the system. Hence we realise an ideal circulator. Using the quantum Langevin equation in the RWA and input-output formalism \cite{2,37,38}, one can write the equations of motion, and hence the Langevin matrix of the system as
\begin{equation}\label{eq17}
\textbf{M} =
\begin{bmatrix}
-\big(i\omega_{1} +\frac{\kappa_{1}}{2} \big)& -ig_{3}e^{-i\phi_{3}} &  -ig_{2}e^{i\phi_{2}} \\
-ig_{3}e^{i\phi_{3}} & -\big(i\omega_{2} +\frac{\kappa_{2}}{2} \big)&  	-ig_{1}e^{-i\phi_{1}} \\
-ig_{3}e^{-i\phi_{2}} &  -ig_{2}e^{i\phi_{1}} &-\big(i\omega_{3} +\frac{\kappa_{3}}{2} \big)\\
\end{bmatrix},
\end{equation}
\noindent
where $\kappa_{i}$ is the decay rate of the $i$-th mode. We see that $\textbf{M}$ is not symmetric about the diagonal. In order to see the non-reciprocal response, one can solve the above system in the Fourier space, obtaining the scattering matrix $\textbf{S}$ that relates the input signal to the mode converted output signal as $\hat{a}_{out} = \textbf{S}\,\hat{a}_{in}$. By tuning the parameters, it is clear from the ratio of scattering amplitudes that symmetry of the system has indeed been broken\textcolor{black}{, as shown in Fig. \ref{fig:sAmp}. Finally, the insertion loss of the system defined as $\textrm{IL}=-10\log_{10}|S_{13}|^2$ has been also computed and plotted in Fig. \ref{fig:iLoss}, showing that it is essentially negligible over the frequency range of interest.} For the simulation done, we have taken $\omega_1=2\pi\times 1\textrm{GHz}$, $\omega_2=2\pi\times 1.05\textrm{GHz}$, $\omega_3=\omega_1+\omega_2$, $g_1=g_2=g_3=2\pi\times 1\textrm{GHz}$, and $\kappa_1=\kappa_2=\kappa_3=2g_1$.

Physically, in the two schemes discussed above, $\Delta\phi$ acts as an artificial gauge flux and plays the role of the magnetic field in a conventional circulator \cite{34}. The lack of an external magnetic bias is advantageous in situations such as qubit read-out schemes, where the performance is severely hindered by the unwanted noises that enter the system through various ports. Hence such a scheme could be a precursor to devices which achieve non-reciprocity between such physical ports of the system.

\section*{Conclusions}
In summary, we have shown that a layered 2D quantum capacitor with graphene electrodes is highly nonlinear, if temperature is held sufficiently low and there is no bias. All requirements were shown to be met at microwave (GHz) frequencies, temperatures of the order of 1K, and capacitors as small as a few $\mu{\textrm{m}}^2$. We showed that under the correct choice of strong pump and weak fields, the interaction Hamiltonian would be that of a beam splitter with adjustable phase from the pump, depending on the resonance condition. Potential applications were discussed and it was shown that the interaction could be easily forced into the ultrastrong regime. It is anticipated that this proposal would open up a new highway in the field of quantum microwave electro-optics, while allowing access to interaction rates which are not easily achievable though other methods. The high strength of Kerr nonlinearity makes the proposed quantum capacitor potentially very useful for design of novel charge quantum bits as well.

\section*{Acknowledgment}
The authors wish to thank Dr. Alexey Feofanov and Dr. Dalziel Wilson at \'{E}cole Polytechnique F\'{e}d\'{e}rale de Lausanne, and Dr. Simon Nigg at the University of Basel for reading the manuscript and making useful suggestions.

\section*{Contributions}
The idea and analysis of 2D quantum capacitor was conceived by S. K. Circulator design and analysis was done by A. K. Both authors discussed the results and contributed significantly to the write up of the paper.

\section*{Competing interests}
The authors declare no competing financial interests.

\section*{Funding}
S. K. is being supported in part by Research Deputy of Sharif University of Technology over a sabbatical visit. This research has been supported in part by the Laboratory of Photonic and Quantum Measurements at \'{E}cole Polytechnique F\'{e}d\'{e}rale de Lausanne.

\end{document}